\documentclass[12 pt, conference,onecolumn]{ieeeconf}  %

\IEEEoverridecommandlockouts                              %

\usepackage{amsmath}
\usepackage{amsfonts}
\usepackage{amssymb} 
\usepackage{graphicx}
\usepackage{subcaption}
\usepackage{xcolor}
\usepackage{chemformula}
\usepackage{bm}
\usepackage[ruled,vlined]{algorithm2e}
\usepackage{hyperref}
\usepackage{mathtools}
\DeclarePairedDelimiter{\abs}{\lvert}{\rvert}
\newcommand\sgn{\mbox{sgn}}
\newtheorem{theorem}{Theorem} 
\title{\LARGE \bf
Mediating Ribosomal Competition by Splitting Pools  
}

\author{Jared Miller, M. Ali Al-Radhawi,  and Eduardo D. Sontag%
\thanks{The authors are with the Department of Electrical \& Computer Engineering, Northeastern University, Boston, MA. E.D. Sontag, is also with the Department of Bioengineering, Northeastern University, and the Laboratory of Systems Pharmacology, Harvard Medical School, Boston, MA. Emails: \texttt{\{miller.jare,malirdwi,e.sontag\}@northeastern.edu}.}}

\graphicspath{{./paper_img/}}

\begin{document}

\maketitle
\thispagestyle{empty}
\pagestyle{empty} %

\begin{abstract}
	Synthetic biology constructs often rely upon the introduction of ``circuit'' genes into host cells, in order to express novel proteins and thus endow the host with a desired behavior. The expression of these new genes ``consumes'' existing resources in the cell, such as ATP, RNA polymerase, amino acids, and ribosomes. Ribosomal competition among strands of mRNA may be described by a system of nonlinear ODEs called the Ribosomal Flow Model (RFM). The competition for resources between host and circuit genes can be ameliorated by splitting the ribosome pool by use of orthogonal ribosomes, where the circuit genes are exclusively translated by mutated ribosomes. 
	In this work, the RFM system is extended to include orthogonal ribosome competition. This Orthogonal Ribosomal Flow Model (ORFM) is proven to be stable through the use of Robust Lyapunov Functions. 
	The optimization problem of maximizing the weighted protein translation rate by adjusting allocation of ribosomal species is formulated and implemented.
\end{abstract}

\section{Introduction}

\label{sec:intro}
The process of protein expression is mediated by ribosomes. After a gene in DNA has been transcribed into messenger RNA (mRNA),  a ribosome binds to the mRNA and begins protein translation. The mRNA is divided into a set of 3-nucleotide segments called codons, and each codon corresponds to an amino acid or a stop instruction. The ribosome attracts a tRNA carrying an amino acid which matches the currently read codon, and appends it to a growing polypeptide chain. Once the ribosome hits a stop codon, it falls off the mRNA and releases the amino acid chain as a polypeptide, which is subsequently post-translationally processed into a final protein product. The mRNAs remain in the cell until they are degraded or destroyed (such as by siRNA or RISC). Intact mRNAs continually attract ribosomes and produce protein, and all mRNAs in the cell compete for a finite pool of resources which includes ribosomes. 

One way to decouple host and circuit genes is to split the pool of resources via orthogonal ribosomes \cite{Rackham2005, Costello2020}. Orthogonal ribosomes are mutated ribosomes which only translate specifically modified circuit genes, and can be constructed by introducing a synthetic 16s rRNA into the cell. The circuit's mRNAs binding sites are designed so that only mutated ribosomes will translate circuit mRNAs; host ribosomes are not attracted to the circuit mRNAs \cite{chubiz2008computational}. 
Orthogonal ribosomes may be able to decrease competition and increase protein throughput by splitting the pool.

We present the Orthogonal Ribosomal Flow Model (ORFM) extending the existing RFM. The ORFM system is a conservative system where the total number of ribosomes is conserved. Global asymptotic stability of the ORFM is certified by means of robust Lyapunov functions. A simple bisection algorithm is detailed to compute the unique ORFM steady state. The protein throughput of the ORFM system can be changed by adjusting the production rate of different species of ribosomes, and maximizing this throughput is a non-concave optimization problem.

The structure of the paper is as follows: 
Section \ref{sec:prelim} reviews existing models for protein translation and ribosome competition. Section \ref{sec:orfm} introduces the ORFM. 
Section \ref{sec:feedback} adds a feedback controller to regulate production of ribosomes. 
Section \ref{sec:optimization} presents a non-concave optimization problem to maximize protein throughput. Section \ref{sec:conclusion} concludes the paper. The stability of the ORFM is proved in the Appendix.

\section{Ribosomal Flow Models (RFMs)}
\label{sec:prelim}

\subsection{Single-Strand Translation Models}

The first models analyzing steady-state ribosomal distributions on mRNA were based on lattice models. In a Totally Asymmetric Simple Exclusion Process (TASEP), mRNA is represented as a finite-length one-dimensional lattice of codons \cite{macdonald1968kinetics, macdonald1969concerning}. Each ribosome waits a random amount of time (exponentially distributed%
) before attempting to move rightwards to the next codon. 

The RFM is a deterministic mean-field approximation to the TASEP model \cite{reuveni2011genome}. Each codon on the mRNA has a normalized ribosomal density $x_j(t)$ $\in$ $[0,1]$, which one may think of as the probability that a ribosome is present on codon $j$ at time $t$. Transition rates between codons 
are denoted here as $\lambda_j$. The initiation rate $\lambda_0$ is the rate at which the mRNA  attracts the ribosome to begin translation, while the $\lambda_{j < n}$ are elongation rates, and represent the amount of time required for the ribosome to attract the codon's respective tRNA. Finally, $\lambda_n$ is the rate at which ribosomes separate from the mRNA and release the completed polypeptide chain. The quantity $y = \lambda_n x_n$ is the rate (protein/time) at which protein is produced by the mRNA. An RFMIO (input/output) is a tridiagonal polynomial single-input single-output system:
\begin{align}
\label{eq:rfmio}
\dot{x}_1 &= \lambda_0 u (1-x_1) - \lambda_1 x_1 (1-x_2)   \\ \nonumber
\dot{x}_{j} &= \lambda_{j-1} x_{j-1} (1-x_j) - \lambda_{j} x_{j} (1-x_{j+1}), 1 < j < n\\ \nonumber
\dot{x}_n &= \lambda_{n-1} x_{n-1} (1-x_n) - \lambda_n x_n \\ \nonumber
y &= \lambda_n x_n  
\end{align}
The output $y$ is the described translation rate, and the input $u$ is the rate at which new ribosomes become available. Ribosomal flux for each codon $x_j$ is based on inflow minus outflow, where the individual flows are determined by parabolic current-density relations multiplied by the elongation rate. Ribosomes are more likely to flow from codon $x_{j}$ to $x_{j+1}$ if $x_{j}$ is full and $x_{j+1}$ is empty, inspiring the term $(1-x_{j}) x_{j+1}$. The state coordinates $x_j$ are occupancy fractions, thus restricted to take values in $[0,1]$

\begin{figure}[t]
	\centering
	\begin{subfigure}[ht]{\linewidth}
		\includegraphics[width=\linewidth]{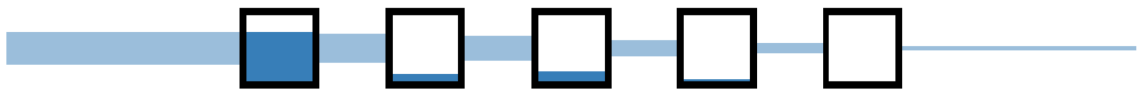}
	\end{subfigure}
	\begin{subfigure}[b]{\linewidth}
		\includegraphics[width=\linewidth]{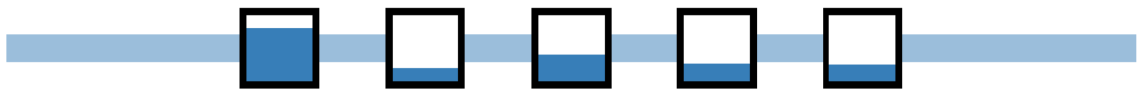}
	\end{subfigure}
	\caption{A pictorial representation of the inflow, outflow, and densities of a 5-codon RFMIO at times $t = 1.5$ and $\infty$.}
	\label{fig:RFM_single}
\end{figure}

For a constant input $u>0$, there is a unique steady-state in $[0,1]^n$ for system \eqref{eq:rfmio}, which we denote as $e$. Figure \ref{fig:RFM_single} shows the steady state of an 5-codon RFM. 
Each codon is a black box, and $x_j$ is the filled proportion of each codon. Ribosomes flow from left to right, and the bars between codons show the flux rates $\lambda_j x_j (1-x_{j+1})$ (which must be all equal, as is clear from the equations). The steady state output $y = \lambda_n e_n$ may be computed by solving a finite continued fraction, which results in a polynomial equation of degree $(n+1)/2$ \cite{reuveni2011genome}. A spectral formulation for the constant $u$ was presented by Poker \emph{et. al.}, and is reviewed in Algorithm \ref{alg:rfmio}. The operator $\textbf{tridiag}(\alpha, \beta)$ for $\alpha \in \mathbb{R}^n, \beta \in \mathbb{R}^{n-1}$ produces a symmetric tridiagonal matrix with main diagonal $\alpha$ and 1-off-diagonals $\beta$. For a square matrix $M$,  $\sigma_{max}(M)$ and $v_{max}(M)$ are the dominant eigenvalue and eigenvectors. By the Perron-Frobenius theorem, $\zeta_1$ has nonnegative entries.

\begin{algorithm}[h]
	\SetAlgoLined
	\SetKwInOut{Input}{input}
	\SetKwInOut{Output}{output}    
	\Input{Rates $\lambda$, Constant Input $u$} 
	\Output{Codon Steady States $e_j$}
	$\mu_j = 1/\sqrt{\lambda_j}$ for $j = 0 \ldots n$\\
	$J = \textbf{tridiag}(\mathbf{0}_{n+2}, \mu)$ \\ 
	$\sigma = \sigma_{max}(J), \ \zeta = v_{max}(J)$ \\
	$e_j = \frac{\mu_j}{\sigma} \frac{\zeta_{j+2}}{\zeta_{j+1}}$ for $j = 1 \ldots n$
	\caption{\label{alg:rfmio} RFMIO Steady State (from \cite{poker2014maximizing})}
\end{algorithm}

The RFM system is a monotone control system, and the set of admissible controls $u \in \mathcal{U}$ is the set of bounded and measurable functions $u \in \mathbb{R}_+$. The RFM is also state and output-controllable, and desired translation rates and patterns can be achieved by proper choice of $u$ and $\lambda$ \cite{zarai2016controllability}.

\subsection{Ribosomal Competition}
\label{sec:rfmnp}
In real genetic systems, protein translation on a strand of mRNA does not occur in isolation. At any particular moment, all mRNA's in the cell compete for a finite (and possibly time-varying) number of ribosomes. 
Particularly slow transition rates $\lambda$ on codons can lead to strands of mRNA hoarding ribosomes, reducing the availability of ribosomes in the pool for all other mRNA's and leading to a globally depressed translation rate. 

Competition was introduced into the RFM framework in the case of one strand of mRNA by positive feedback \cite{margaliot2013ribosome} and on circular mRNA \cite{raveh2015ribosome}, in each case resulting in systems with a globally asymptotically stable steady state. Raveh \emph{et. al.} introduced a ribosomal flow model network with a pool (RFMNP) to abstractly describe the impact of ribosomal competition \cite{raveh2016model}. Each of the $s$ strands of mRNA is modeled by a RFMIO ($x^i_j$ for codon $j$ of mRNA $i$), and all RFMIO are connected to a common pool $z$. The total number of ribosomes in the system $N_r = z(t) + \sum_{i, j} x^i_j (t)$ is a conserved quantity. The input $u$ of each mRNA is a monotonically increasing concave saturation function $G^i(z)$ (commonly $z$ or $\text{tanh}(z)$), which describes the likelihood that a ribosome from the pool will attach itself to strand $i$. The output $y^i = \lambda_n^i x_n^i$ is the translation rate, and ribosomes leaving $x_i$ return to the pool. The pool dynamics are therefore: \vspace{-0.15in}

\begin{equation}
\dot{z} = \textstyle\sum_i{\lambda^i_n x^i_n} - \textstyle\sum_i{\lambda^i_0 G(z)(1- x^i_1)}
\end{equation}

\vspace{-0.04in}

The RFMNP is a closed loop system. The number of ribosomes $N_r$ defines a stoichiometric class, and RFMNP has a globally asymptotic equilbrium point with all $e_j^i \in (0, 1)$ and $z \in (0, N_r)$ for each $N_r$. Stability can be proven by contraction, where the weighted $L_1$ norm between trajectories is non-expanding over time \cite{raveh2016model}.

\begin{figure}[ht]
	\centering
	\begin{subfigure}[b]{0.48\linewidth}
		\includegraphics[width = \linewidth]{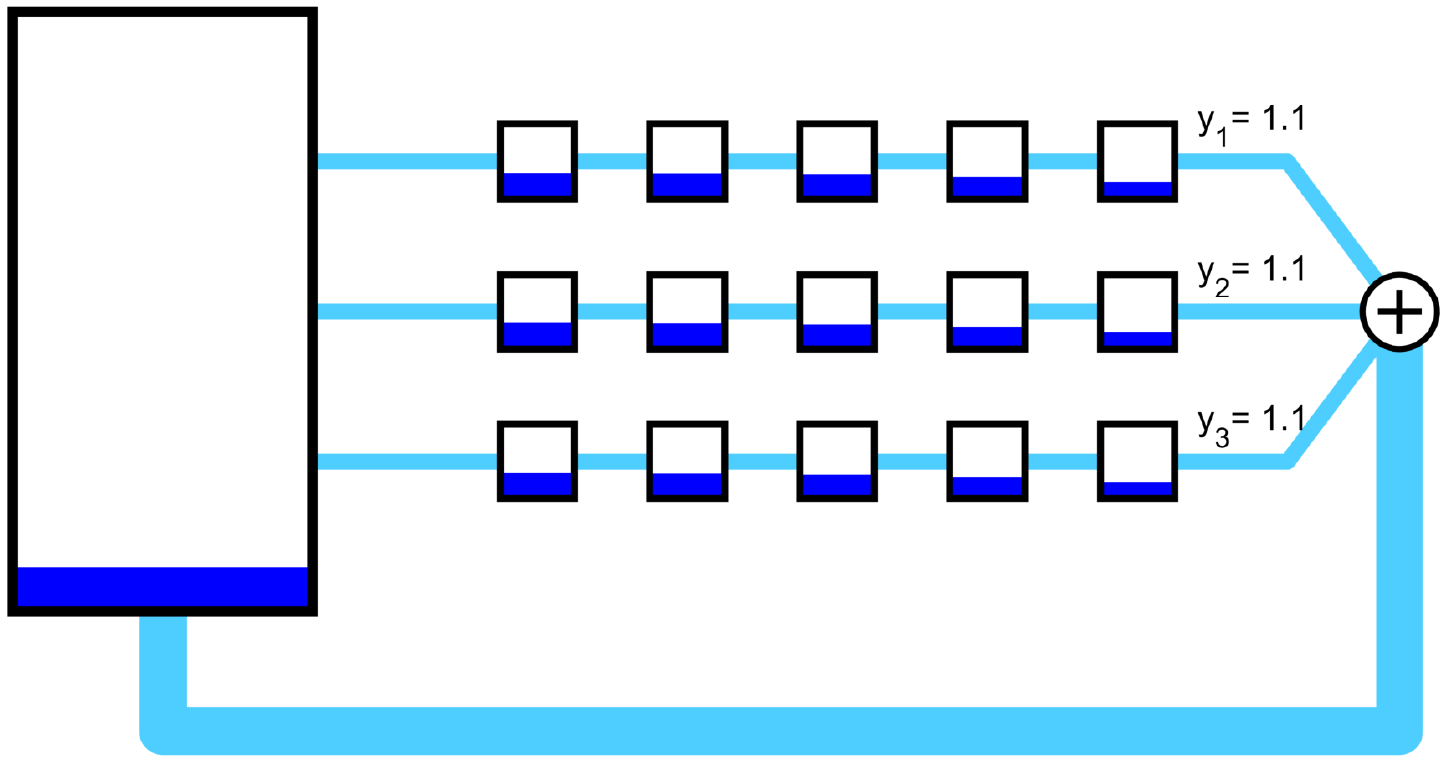}
		\caption{Homogeneous transitions.}
		\label{fig:RFMNP_homog}
	\end{subfigure}
	\begin{subfigure}[b]{0.48\linewidth}
		\includegraphics[width = \linewidth]{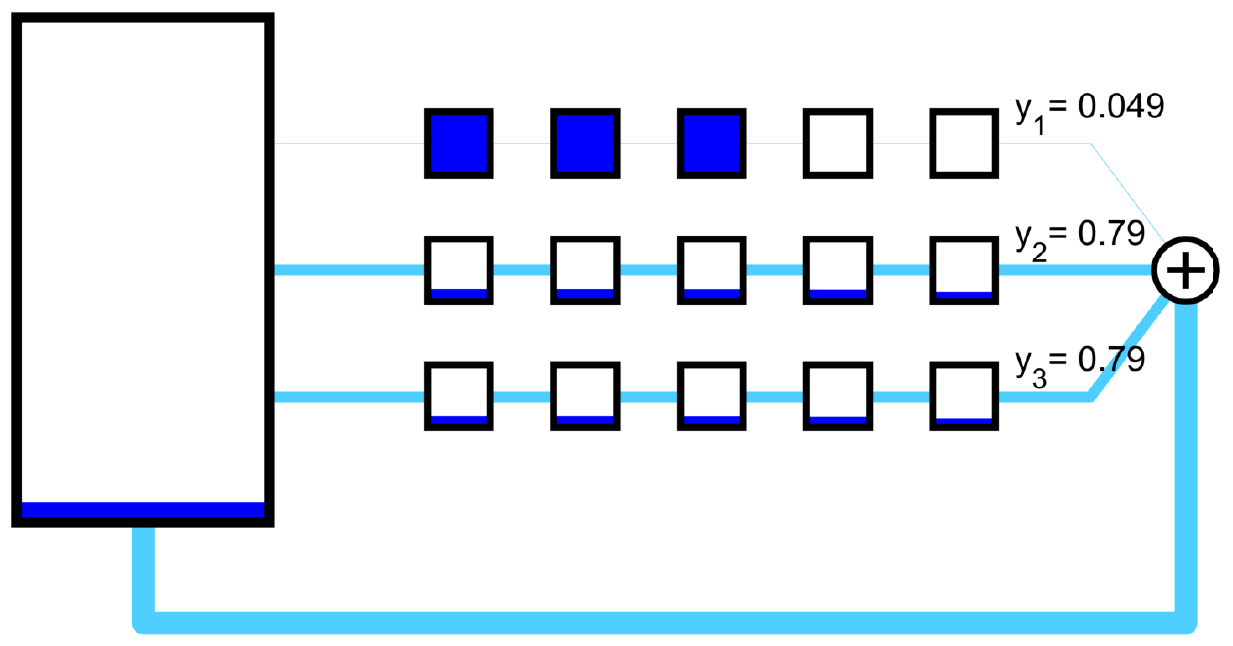}
		\caption{Strand 1 is slow}
		\label{fig:RFMNP_bottleneck}
	\end{subfigure}
	
	\begin{subfigure}[b]{0.48\linewidth}
		\includegraphics[width = \linewidth]{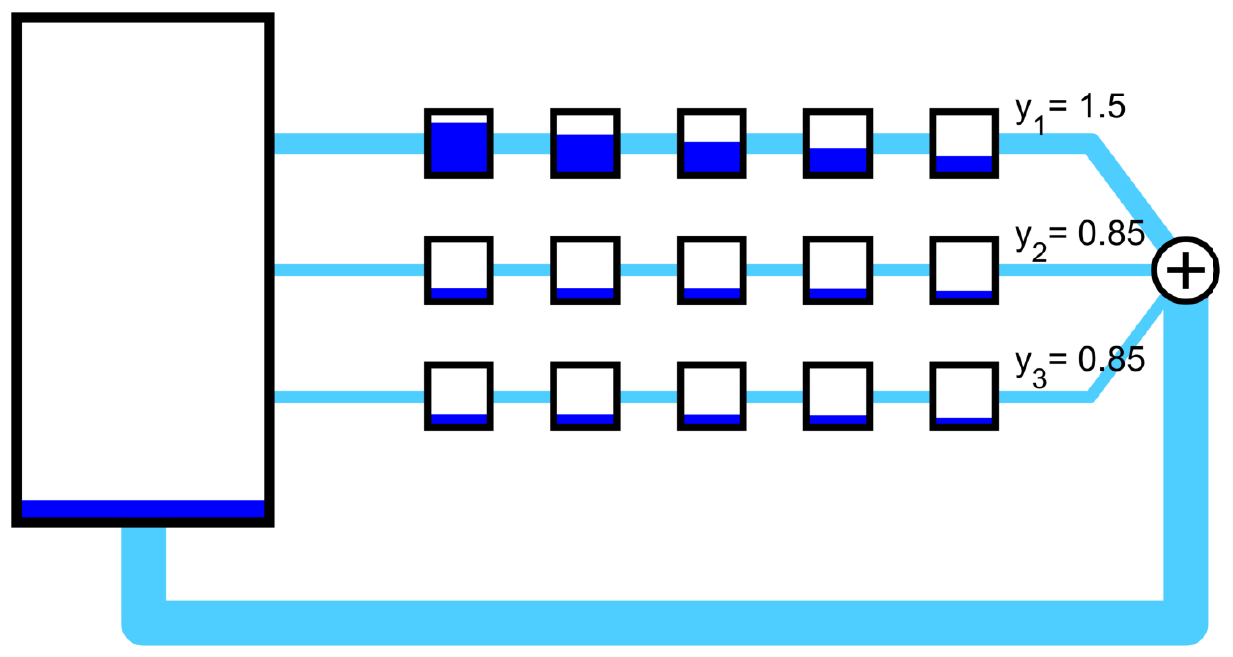}
		\caption{Strand 1 is greedy}
		\label{fig:RFMNP_greedy}
	\end{subfigure}
	\begin{subfigure}[b]{0.48\linewidth}
		\includegraphics[width = \linewidth]{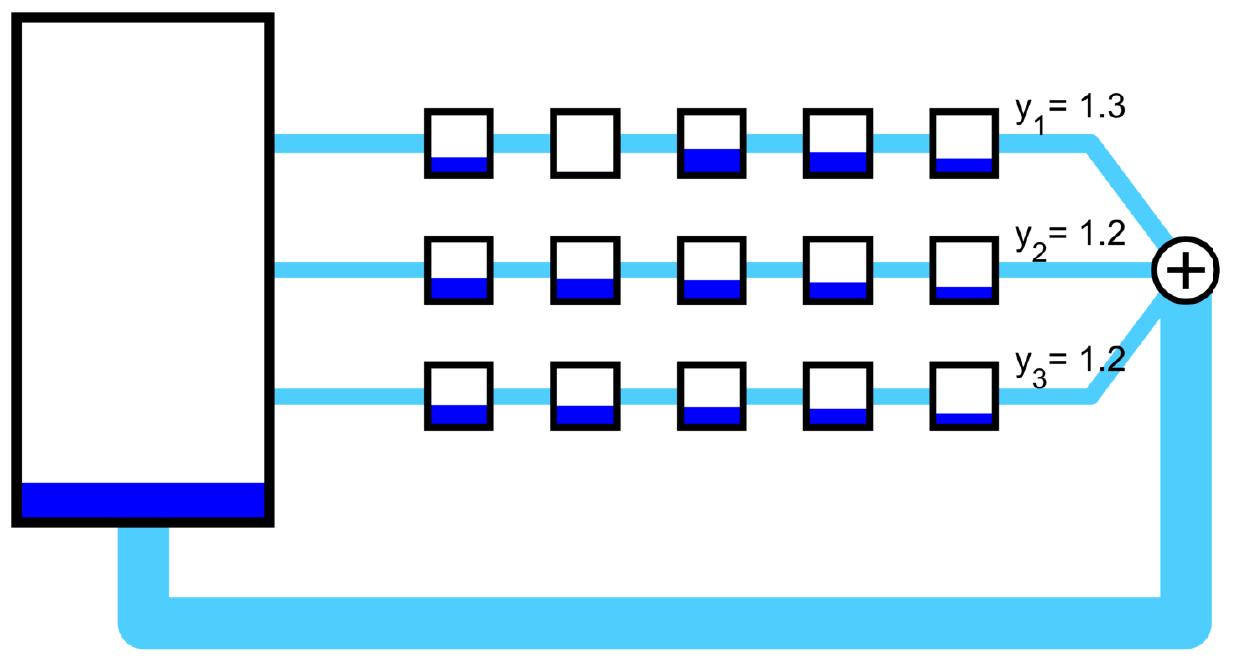}
		\caption{Strand 1 is fast}
		\label{fig:RFMNP_speedy}
	\end{subfigure}
	\caption{\label{fig:RFMNP} RFM competition at steady state}
	\vspace{-0.15in}
\end{figure}

Competition effects can be observed by perturbing parameters $\lambda^i_j$ and analyzing resultant steady states. If $\lambda^i_j$ changes for a specific strand of mRNA $x^i$, then the protein translation rate $y^i$ will change in that same direction (increasing $\lambda^i_j$ will increase $y_j$). The translation rate of all other mRNA in the network will change uniformly: an increase in $\lambda^i_j$ will raise $y^i$ and will either raise or lower all $y^{k \neq i}$. Further discussion on competition effects can be found in Section 4.2 of \cite{raveh2016model}, and some of these results are demonstrated in Figure \ref{fig:RFMNP}.
Figure \ref{fig:RFMNP} shows four examples of an RFMNP where each of the three strands of mRNA have five codons each. All strands start with homogeneous transition rates of $\lambda^i_j = 5$, and there are a total of $N_r = 5$ ribosomes in the system. Figure \ref{fig:RFMNP_homog} visualizes the steady state of this homogenous system, where each mRNA has a protein translation rate of  $y^{1:3}_{\mathit{ss}} = 1.15$. Ribosomes cannot easily pass from codons 4 to 5 because $\lambda^1_3 = 0.05$. This bottleneck drops the translation efficiency of strand 1 to $y^1_{\mathit{ss}} = 0.049$ and the others to $y^{2:3}_{\mathit{ss}} = 0.79$. Figure \ref{fig:RFMNP_greedy} has $\lambda^1_0 = 40$, so strand 1 attracts ribosomes from the pool at a much faster rate than the other strands. The first codon cannot drain ribosomes quickly since $\lambda^1_{1 \ldots  5} = 5$, so ribosomes are stuck in strand 1. Translation rates rise for strand 1 to $y^1_{\mathit{ss}} = 1.50$, and fall on other strands to $y^{2:3}_{\mathit{ss}} = 0.85$. Figure \ref{fig:RFMNP_speedy} modifies strand 1 to  $\lambda^1_{2} = 40$. Ribosomes quickly exit strand 1 and enter the pool again for use in further translation. This improves the translation efficiency of all strands, raising $y^1_{\mathit{ss}} = 1.33$ and $y^{2:3}_{\mathit{ss}} = 1.18$.

\section{Orthogonal Ribosomal Flow Model}
\label{sec:orfm}
Orthogonal ribosomes can be added to the RFM scheme (ORFM) by splitting the ribosome pool $z$. 
Code for simulation and visualization is publicly available online at  
\url{https://gitlab.com/jarmill/Ribosomes}.

\subsection{ORFM Formulation}

The proposed model of ribosome translation has $M$ species of ribosomes. Each ribosome species has a pool $z_p$ for $p = 1 \ldots M$ of available ribosomes. Strands of mRNA that use ribosome type $p$ form an RFNMP with pool $z_p$.

Ribosomes of type $p$ are formed by the combination of a os-15 rRNA of type $p$ and the remaining ribosome components. The protein backbone and large ribosomal subunit are treated as an `Empty' ribosome $E$, which has no translation capacity on its own. 
Empty ribosomes bind with rRNA at rate $k^+_p$, and the ribosomal complex dissociates at rate $k^-_p$, assuming an effectively infinite amount of rRNA available. These kinetics are inspired by the work of Darlington \emph{et. al.} in their mass-action model of protein translation and cell metabolism \cite{darlington2018dynamic}. The $M$ pools of translating ribosomes $z_p$ are each connected to the pool of empty ribosomes $z_E$.

Each of the $s_p$ strands of mRNA that obtain ribosomes from pool $z_p$ will have a corresponding length $n^{pi}$ and are repeated with multiplicity $m^{pi}$. The codon at location $j \in 0 \ldots n^{pi}-1$ of RNA strand $i \in 1 \ldots s_p$ coming from pool $p$ 
is $x_j^{pi}$. The constants of this system are the $N_r$ ribosomes, binding rates $k_{p}^+$ and $k_{p}^-$, and transition rates $\lambda_j^{pi}$. 
The total number of ribosomes $N_r$ is a conserved quantity:
\begin{equation}
\label{eq:general_conservation}
N_r = z_E + \textstyle\sum_{p=1}^M{ z_p} + \textstyle\sum_{p=1}^M{  \textstyle\sum_{i=1}^{s_p}{\left(m^{pi}\textstyle\sum_{j=1}^{n^{pi}} {x^{pi}_j}\right)}} \end{equation}
The translation dynamics are:
\begin{subequations}
	\begin{align}
	\label{eq:general_model}
	\dot{z}_E &= \textstyle\sum_p {k_{p}^- z_p} - \textstyle\sum_p{k_{p}^+} z_E  \\
	\dot{z}_p &= k_{p}^+ z_E - k_{p}^- z_p  + \textstyle\sum_i{ m^{pi} \lambda_{n^{pi}}^{pi} x_{n^{pi}}^{pi}} \nonumber\\ 
	& \quad - \textstyle\sum_i{m^{pi} \lambda_0^{pi} (1-x_1^{pi})  G^{pi}(z_p)} \\
	\dot{x}_1^{pi} &= \lambda_0^{pi}  (1-x_1^{pi}) G^{pi}(z_p) -\lambda_1^{pi}  (1-x_2^{pi}) x_1^{pi} \label{sys_G}\\
	\dot{x}_j^{pi} &= \lambda_{j-1}^{pi}  (1-x_{j}^{pi}) x_{j-1}^{pi} -\lambda_j^{pi}  (1-x_{j+1}^{pi}) x_j^{pi} \\
	\dot{x}_{n^{pi}}^{pi} &= \lambda_{n^{pi}-1}^{pi}  (1-x_{{n^{pi}}-1}^{pi}) x_{{n^{pi}}}^{pi} -\lambda_{n^{pi}}^{pi}    x_{n^{pi}}^{pi}  \label{eq:general_model_}
	\end{align}
\end{subequations}

Define $\bar{N}_r = z_E + \sum_{p} z_p$. For a fixed $\bar{N}_r$, the number of ribosomes in each pool can be obtained by solving a linear system where $K_p = k_p^+ / k_p^-$:
\begin{equation}
\label{eq:pool_sys}
z_p= K_p z_E \qquad \bar{N}_r = z_E + \textstyle\sum_p z_p
\end{equation}
\begin{algorithm}[h]
	\SetAlgoLined
	\SetKwInOut{Input}{input}
	\SetKwInOut{Output}{output}    
	\Input{$\lambda_j^{pi}, \ N_r, \ G^{pi}, \ K_p, \ \epsilon $} 
	\Output{Steady States $z_E, z_p, e_j^{pi}$}
	$\bar{N}_{min} = 0, \ \bar{N}_{max} = N_r$\\
	\Repeat{$\abs{N_{curr} - N_r} \leq \epsilon$}{
		$\bar{N}_{curr} = (\bar{N}_{max} + \bar{N}_{min})/2$\\
		$z_E = \frac{\bar{N}_r}{1 + \sum_p K_p} \qquad z_p = \frac{K_p\bar{N}_r}{1 + \sum_p K_p}$ \\
		$e^{pi} = \text{RFMIO SS}(\lambda^{pi}, G^{pi}(z_p))$   (Alg. \ref{alg:rfmio})\\
		$N_{curr} = \bar{N}_{curr}+ \sum_{p, i, j} e^{pi}_j $ \\
		\eIf{$N_{curr} \leq N_r$}{
			$z_{min} \leftarrow z_{curr}$
		}{
			$z_{max} \leftarrow z_{curr}$
		}
	}
	\caption{\label{alg:rfm_general} ORFM Steady State}
\end{algorithm}
Algorithm \ref{alg:rfm_general} computes the steady-state of the general model by bisection on $\bar{N}$. This algorithm can compute the steady state of an RFMNP by treating the network as a $1$-species model with $K_1 \rightarrow \inf$ where $\bar{N} = z$.

\begin{figure}[t]
	\centering
	\begin{subfigure}[b]{0.45\linewidth}
		\includegraphics[width = \textwidth]{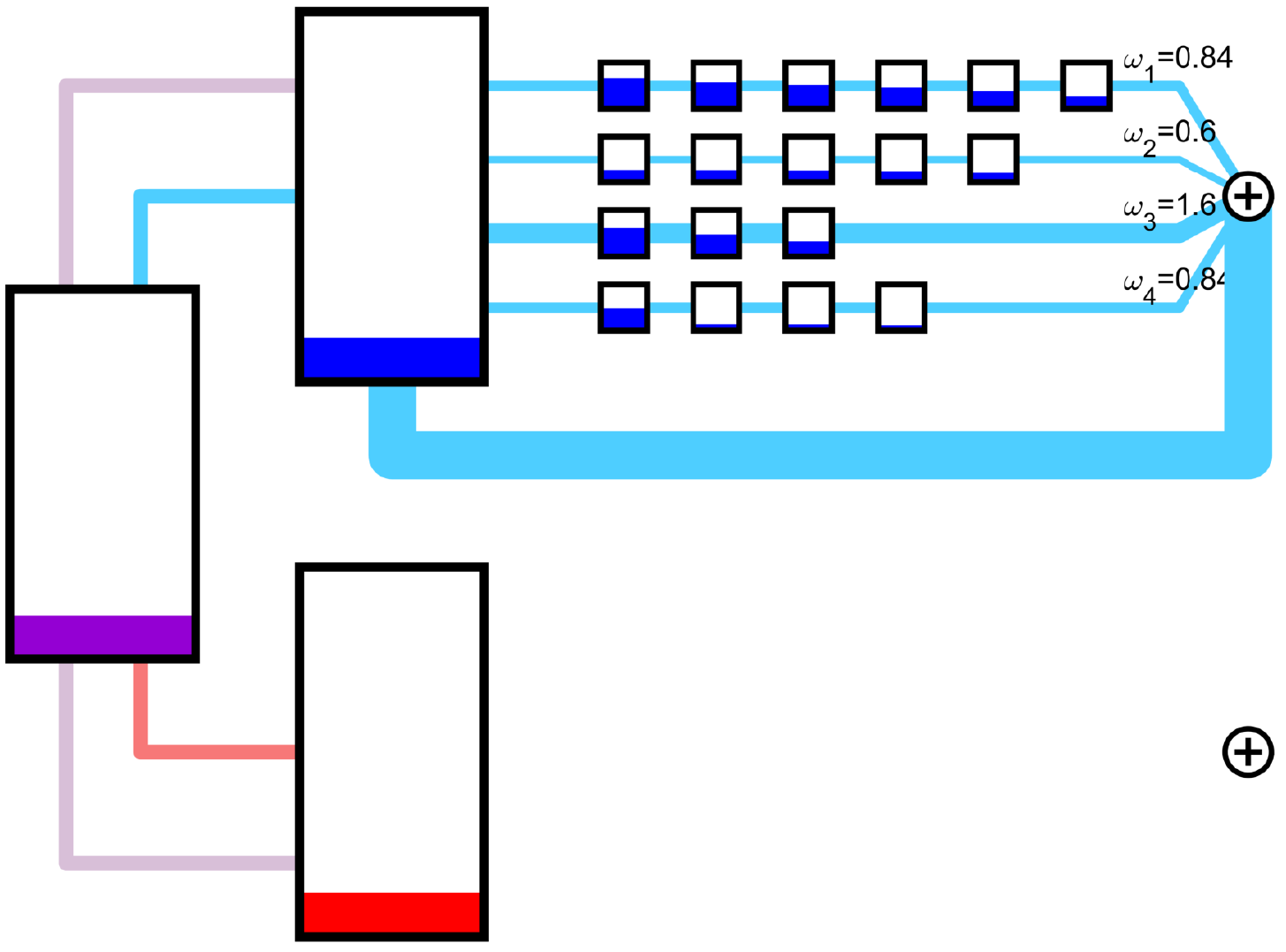}
		\caption{No circuit demand}
		\label{fig:ORFM_no_circuit}
	\end{subfigure} \quad 
	\begin{subfigure}[b]{0.45\linewidth}
		\includegraphics[width = \textwidth]{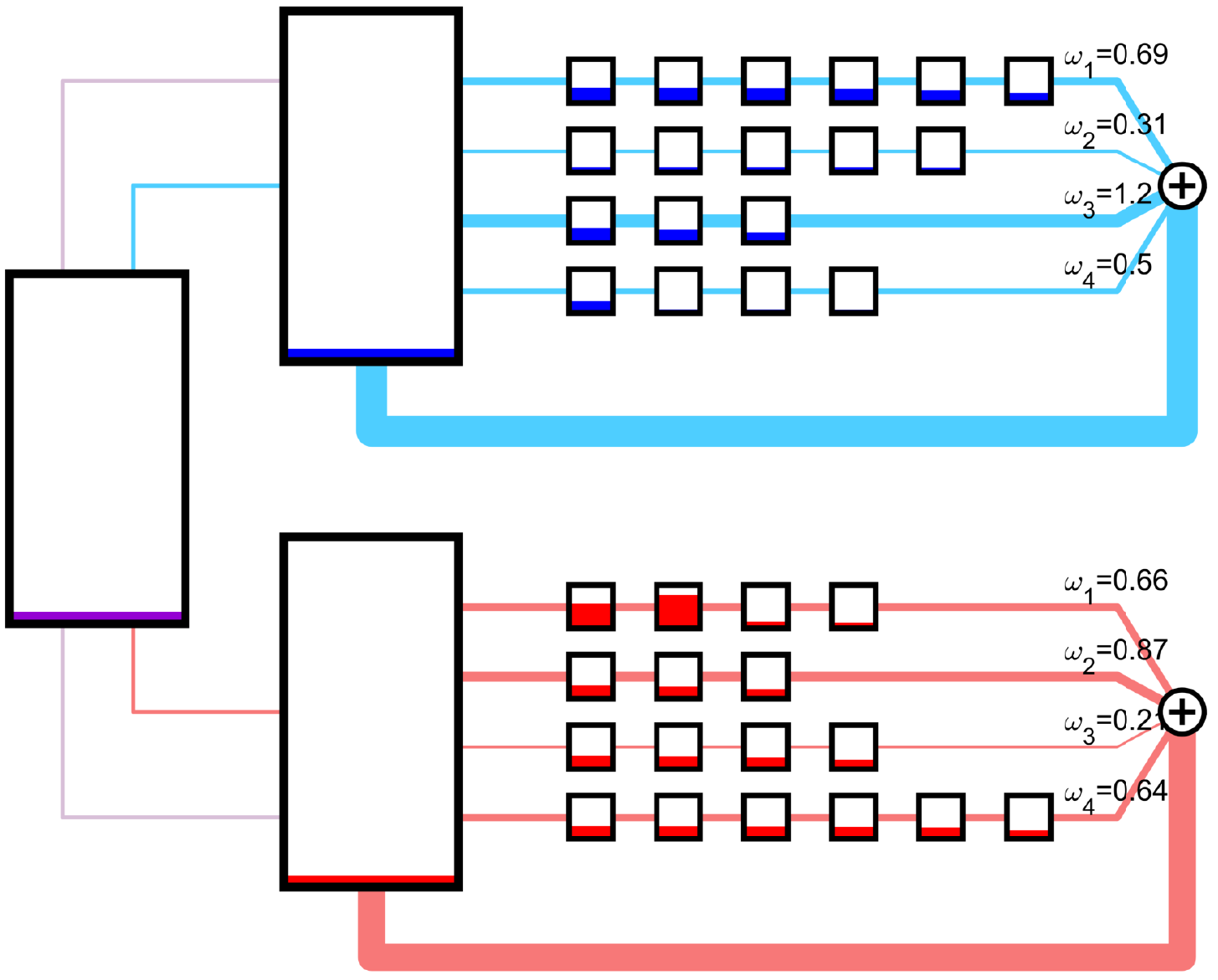}
		\caption{High circuit demand}
		\label{fig:ORFM_high_circuit}
	\end{subfigure}
	\caption{\label{fig:ORFM} Orthogonal RFM visualizations}
\end{figure}

Figure \ref{fig:ORFM} shows examples of mRNA competing for $N_r=10$ ribosomes, where each mRNA may take ribosomes from either the host pool ($p= 1$,  blue) or the circuit pool ($p = 2$, red). The `empty' ribosomes are displayed in purple in the third tank.  In this network, $k_{p}^+ = k_{p}^- = 1$, so at steady state, the pool quantities $z_1 = z_2 = z_E$. In figure \ref{fig:ORFM_no_circuit} no mRNA takes ribosomes from the circuit pool $z_2$, so the ribosomes in $z_2$ induce deadweight loss for the system.

\begin{figure}[t]
	\centering
	\includegraphics[width =0.5\linewidth]{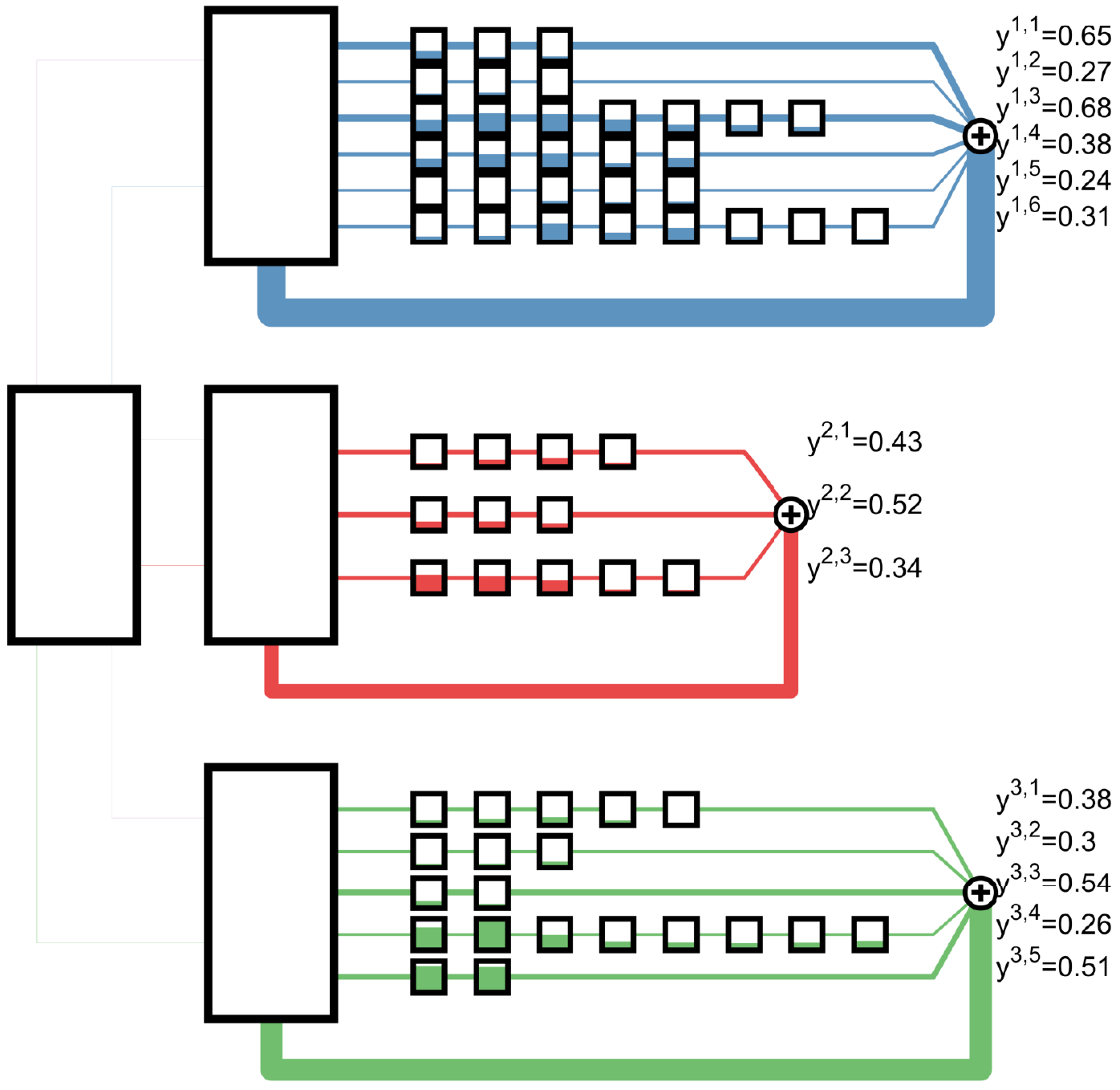}
	\caption{\label{fig:general_tank} ORFM with 3 ribosomal subspecies}
\end{figure}

Figure \ref{fig:general_tank} illustrates a general model orthogonal ribosomal system with pool across three ribosomal subspecies and $N_r$ = 20. 
Now $k_p^+ $$=$$ k_p^- $$= 0.1, p = 1,2,3$, so all pools will have an equal number of ribosomes at equilibrium $(z$$ = $$0.286)$.

\subsection{Stability of the ORFM}
The RFM has recently been studied by constructing explicit Robust Lyapunov Functions (RLFs) \cite{MA_LEARN} via writing it as a Chemical Reaction Network (CRN) and utilizing relevant methods \cite{MA_cdc13,blanchini14,MA_TAC}. Such techniques provide explicit formulae of Lyapunov functions for general kinetics (not limited to Mass-Action kinetics), and have an easy-to-use software package. 

In this subsection, we derive the stability of the ORFM model \eqref{eq:general_model}-\eqref{eq:general_model_} via an RLF. Let $x=:[x_1^{1,1},..,x_{n^{M,s_M}}^{M,s_M}] \in [0,1]^{N_c}$ be the vector of all codon occupancies, and  $N_c$ be the total number of codons. Let $z:=[z_1,..,z_M,z_E]^T\in \mathbb [0,N_r]^{M+1}$ be the vector of all pool occupancies. %
We therefore have: 
\begin{theorem}\label{t.stability}
	Consider the system \eqref{eq:general_model}-\eqref{eq:general_model_}. Then,
	
	\noindent (1) 
	The function:
	$$ V(x,z)= \sum_{p,i} m^{pi} \sum_j |\dot x_{j}^{pi}|+\sum_p |\dot  z_p| + |\dot  z_E|, $$
	is a (non-strict) Lyapunov function for any choice of  $\{\lambda_{j}^{p,i}\}>0$ and monotone functions $G^{pi}$, %
	and 
	
	\noindent
	(2)  %
	For any fixed total number of ribosomes in the system $N_r>0$, there exists a unique positive globally  asymptotically stable steady-state $(x_e,z_e)$ for \eqref{eq:general_model}-\eqref{eq:general_model_}. 

\end{theorem}

A proof can be achieved by transforming \eqref{eq:general_model}-\eqref{eq:general_model_} into a CRN and extracting a stoichiometry matrix. For a fixed number of pools, strands, and codons, Theorem \ref{t.stability} can be verified via the software package \texttt{LEARN} \cite{MA_LEARN}. A proof  for the general result is given in the Appendix.

\subsection{Cross-talk Model}
\label{sec:crosstalk} 
The analysis of the previous sections assume that only host ribosomes translate host genes and only circuit ribosomes translate circuit genes. While it is energetically unfavorable for host ribosomes to translate circuit genes, this event may still happen at a low proability. The phenomenon of a ribosome of species $p$ translating an mRNA that prefers ribosomes of species $p' \neq p$ is a 'cross-talk'. 
Each codon of mRNA still can only accomodate one ribosome at a time, so ribosomes of different species are additionally competing for space. $x_j^p$ is the probability that a ribosome of species $p$ sits at codon $j$, and $x_j = \sum_p x_j^p$ is the probability that some ribosome is present at codon $j$. A cross-talk RFM can be treated as a MISO polynomial (non-tridiagonal) system, with inputs $u^p$ and output $y = \sum_{p' = 1}^M \lambda^p_n x^p_n$:

\begin{align}
\label{eq:crosstalk_open}
\dot{x}_0^{p} &= \lambda_0^{p}  u^p (1-\sum_{p'=1}^{M}{x_0^{p'  i}}) -  \lambda_1^{p} (1-\sum_{p'=1}^{M}{x_{0}^{p}}) x_1^{p} \nonumber\\
\dot{x}_j^{p} &= \lambda_{j-1}^{p}  (1-\sum_{p'=1}^{M}{x_{j-1}^{p'  i}})x_j^{p} -  \lambda_j^{p} (1-\sum_{p'=1}^{M}{x_{j}^{p}}) x_{j+1}^{p} \nonumber\\
\dot{x}_{n^{p}}^{p} &= \lambda_{n^{p}-1}^{p} (1-\sum_{p'=1}^{M}{x_{{n-1}}^{p}}) x_{n^p}^{p} - \lambda_{n}^{p}  x_{n}^{p} \end{align}

\begin{figure}[h]
	\centering
	\caption{\label{fig:cross_rfm_single} mRNA with 3 ribosomal species}
	\includegraphics[width = 0.7\linewidth]{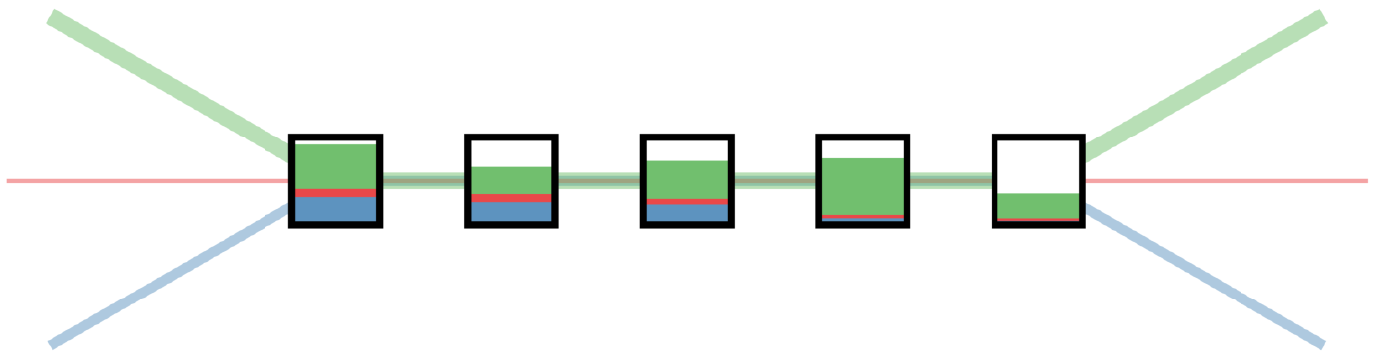}
\end{figure}

There is no requirement that the different transition rates $\lambda^{pi}_j$ be the same between species. Figure \ref{fig:cross_rfm_single} shows a 5-codon mRNA with $\lambda^{p}_j$ uniformly drawn from the range $[0.5, 2.5]$.

A cross-talk RFM where of length $n$ where $\lambda^{p}_j = \lambda^j, j\neq 0$ is equivalent to a standard RFM in terms of sums $s_j = \sum_{p=1}^M {x^p_j}$, with effective initial rate $\lambda^s_0 = \sum_{p=1}^M \lambda^{pi}_0$. Ribosome flux in codons can be equivalently expressed as:

\begin{equation}
\dot{x}^p_j = \lambda_{j-1}(1-s_j)x_{j-1}^p - \lambda_j(1-s_{j+1}) x_j^p
\end{equation}

Equilibrium values for $s$ can be found through spectral methods, and the steady state per-species codon values are:

\begin{equation}
x^{p}_{j<n} = \frac{\lambda^{p}_0}{\lambda_j} \frac{1 -  s_1}{1 - s_{j+1}} \qquad 
x^p_n = \frac{\lambda^{p}_0}{\lambda_n} (1 -  s_1) 
\end{equation}

Stability and a simple method for computing steady-state solutions to the inhomogenous cross-talk RFM are still open questions. In all experiments, the cross-talk RFM is globally asymptotically stable in $x^p_j \in [0, 1]$.

\subsection{Cross-talk Competition}

In an open loop crosstalk model, the various $z_p$ are constant or are external inputs to the system. The crosstalk model can also have network competition, in which the empty-tank pool dynamics are added. As before, $x_j^{pi}$ is the probability that a ribosome of species $p$ is located on codon $j$ of mRNA $i$ at any particular time. The input $u^{pi}$ applied to each strand is $G^{pi}(z_p)$, and pool dynamics $\dot{z}_p$ are adjusted to deal with the cross-talk effects (1-sum occupancy).

\begin{figure}[t]
	\centering
	\includegraphics[width = 0.5\linewidth]{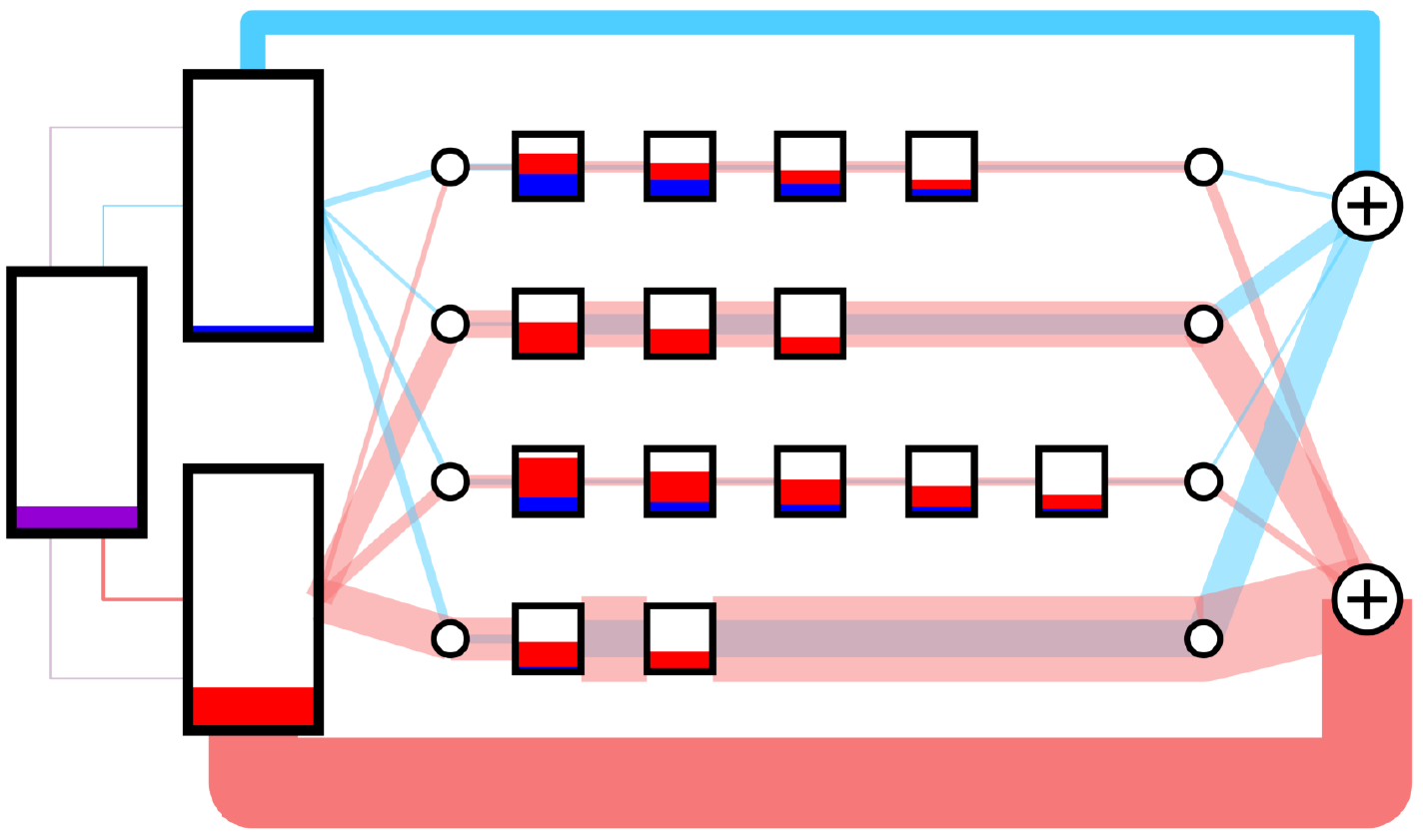}
		\caption{A network of cross-talk mRNA}
			\label{fig:crosstalk_tank}
\end{figure}

Figure \ref{fig:crosstalk_tank} illustrates a cross-talk orthogonal RFM with $N_r = 10$ ribosomes.

\section{Self-Inhibiting Feedback Controller}
\label{sec:feedback}

This section introduces a feedback controller inspired by \cite{darlington2018dynamic} to dynamically generate orthogonal ribosomes as needed in an ORFM. Figure \ref{fig:ORFM_no_circuit} shows an example where there are no mRNAs that take ribosomes from the circuit pool $z_2$. A high $k^+_2$ therefore leads to the creation and maintenance of circuit ribosomes that are not used in translation. A feedback controller can be introduced to create orthogonal ribosomes only as needed, and decrease wastage in the system. This can be accomplished by creating one distinguished mRNA $x^{pF}$ that translates a protein $F_p$ for each non-Host pool. The dynamics of the protein $F_p$ with a degradation rate $\delta_p$ is:
\begin{equation}
\dot{F}_p = \lambda^{pF}_n x^{pF}_n - \delta^p F_p
\end{equation}
A Hill-like inhibition term can be used to suppress the creation of new ribosomes (based on Equation 18 of Supplement 1 of \cite{darlington2018dynamic}). For a nominal ribosome creation rate $k^{+0}_p$, exponent $\gamma_p$, and constant $F_p^0$, the suppressed $k^+_p$ is:
$k^+_p = k^{+0}_p /(1 + F_p/F_{p}^0 )^{\gamma_p}$. 
If $\delta=0$, $F_p$ never degrades once translated. Asymptotically, $k^+_p \rightarrow 0$ and $z_p \rightarrow 0$. A choice of $\delta>0$ will lead to $z_p > 0$, but proper choice in inhibitor parameters can allow for the nonzero $z_p$ to be small if there is no mRNA demand.

\begin{figure}[t]
	\centering
	\begin{subfigure}[b]{0.48\linewidth}
		\includegraphics[width = \textwidth]{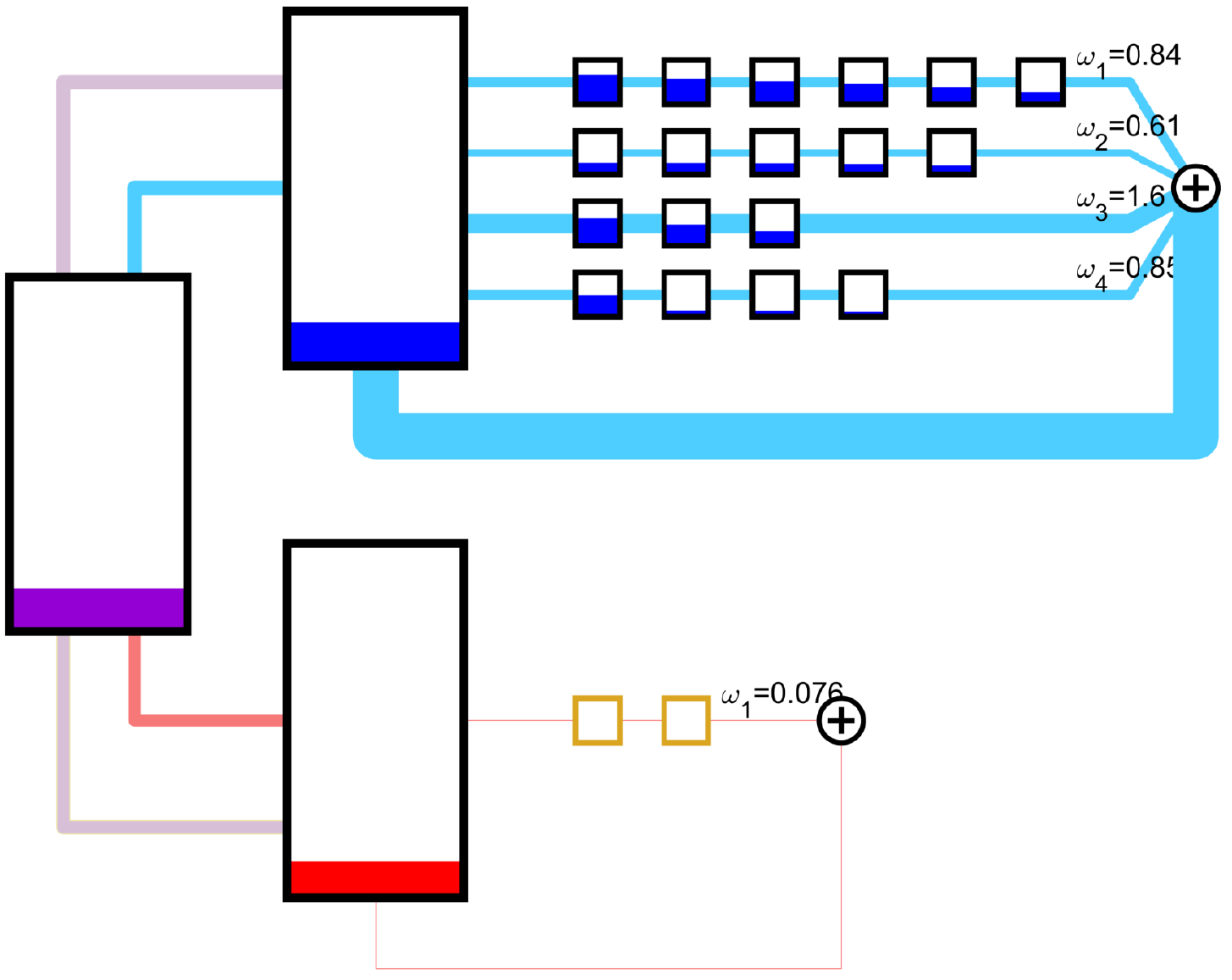}
		\caption{No circuit demand}
		\label{fig:ORFM_inhibitor_no_demand}
	\end{subfigure}
	\begin{subfigure}[b]{0.48\linewidth}
		\includegraphics[width = \textwidth]{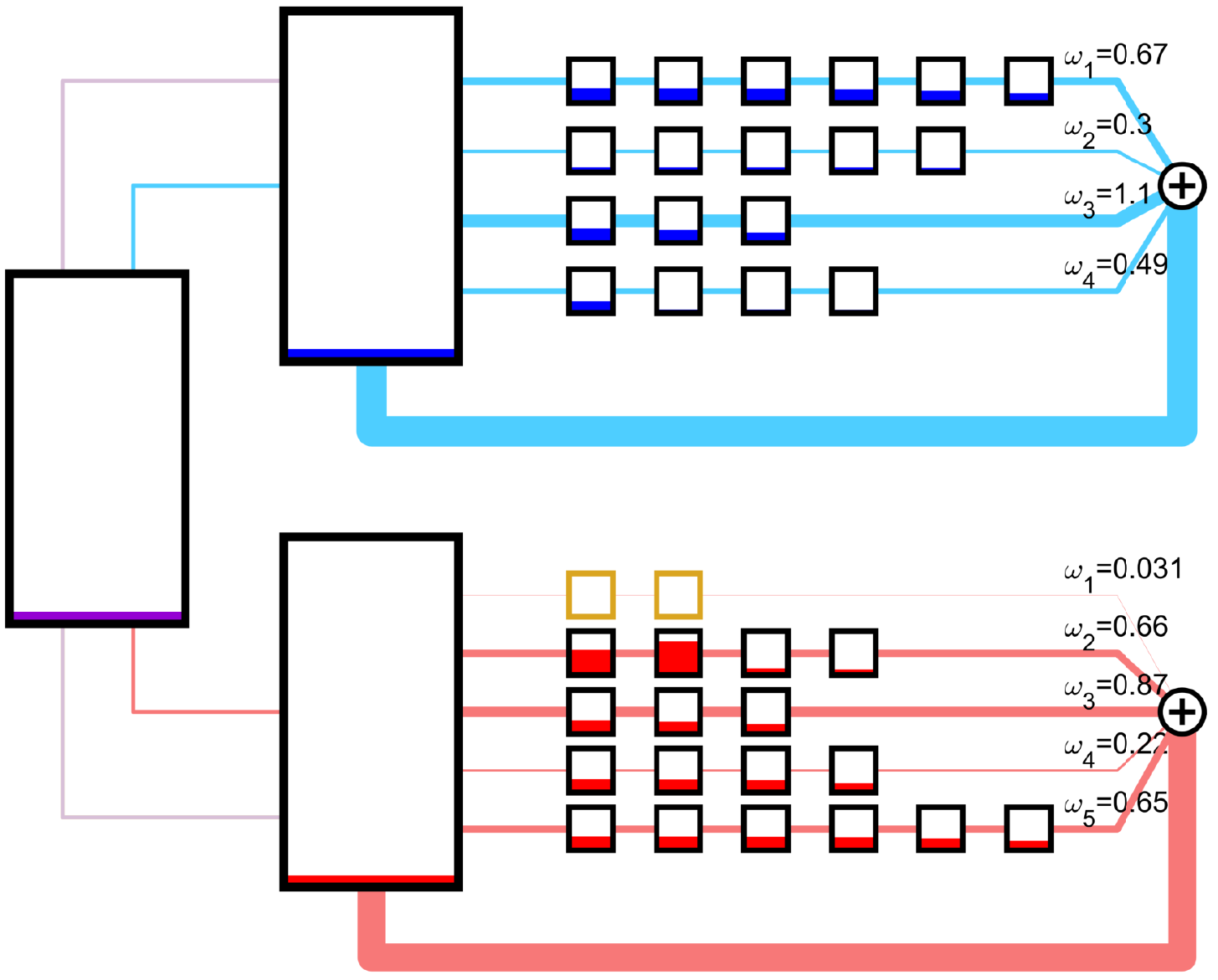}
		\caption{High circuit demand}
		\label{fig:ORFM_inhibitor_high_demand}
	\end{subfigure}
	\caption{\label{fig:feedback}ORFM with inhibitor protein $F_p$.}
\end{figure}

Figure \ref{fig:feedback} shows an example of an ORFM with 10 ribosomes, where the circuit pool $z_c$ has a feedback controller. The golden two-codon mRNA is the distinguished $x^F$ that produces the inhibitor protein $F$ with $\lambda^F = [0.1, 5, 5]$. The comparatively low initiation rate $\lambda^F_0$ allows other mRNA if present to take ribosomes from $z_C$ first. The inhibitor parameters are $F_p^0 = 4, \gamma_p = 2$, and $\delta^p = 0.2$. Simulations are run for $t=0\ldots 300$.

With no inhibitor and no circuit mRNA, the system in Figure \ref{fig:ORFM_no_circuit} has a total of $N_r^H = 7.63$ ribosomes in the host pool and mRNA. With the inhibitor in Figure \ref{fig:ORFM_inhibitor_no_demand}, the total number of host ribosomes rises to $N_r^H = 7.73$. Figure \ref{fig:ORFM_high_circuit} has a high circuit demand, and $N_r^H = 3.94$. With an inhibitor in Figure \ref{fig:ORFM_inhibitor_high_demand}, $N_r^H = 4.05$. At no demand, the steady state $F_p = 0.38$, and at high demand, $F_p = 0.16$.

\section{Optimization}
\label{sec:optimization}
The goal of optimization in this section is to maximize the rate of protein production by mRNA. Optimization of protein throughput in a single strand has previously been treated by Poker \emph{et. al.} in \cite{poker2014maximizing}. Given an $n$-codon mRNA with allowable choices of $\lambda_j$ in a convex set, finding a $\lambda^*$ that maximizes the throughput $y = \lambda_n e_n$ is a concave optimization problem. 

Ribosomes with a pool have an additional layer of feedback. For a single strand with $\lambda_j$ taking ribosomes from pool $z$, the effective intake rate as in Equation \eqref{eq:rfmio} is $\lambda_0 G(z)$. Since $G(z)$ is monotonically increasing and concave, the throughput $y$ is also a concave function of $z$. 

\begin{figure}[t]
	\centering
	\includegraphics[width=0.7\linewidth]{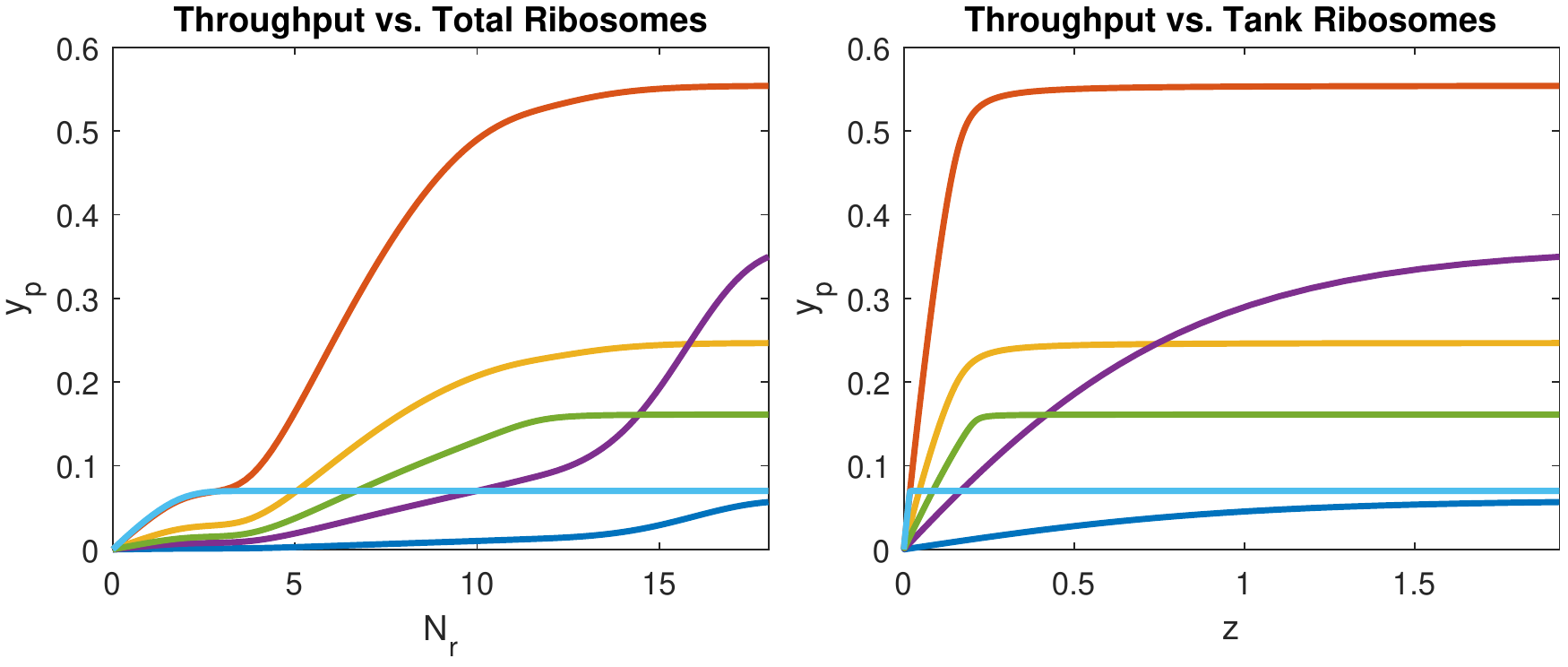}  
	\caption{Protein throughput per mRNA}
	\label{fig:output_tank}
\end{figure}

This concavity breaks when multiple strands of mRNA are connected to the same pool, as shown in Figure \ref{fig:output_tank} for an RFMNP with 6 strands (Section \ref{sec:rfmnp}). The left plot shows the protein production $y^i$ rate of each strand for $N_r \in [0, 18]$. All curves are monotonically increasing with inflection points. This nonconcavity comes from competition effects: an exceptionally greedy strand of mRNA will take ribosomes until it is saturated, and then other strands of mRNA can translate. In contrast, the right hand plot illustrates the translation rate as a concave function of tank capacity. The mapping $N_r \rightarrow z$ carries the competition effects and nonconcavity. It is conjectured that the mapping $N_r \rightarrow y$ is concave if there is only a single strand connected to the pool.

The objective in this section is to maximize the weighted sum of protein production $y_w = \sum_{p,i} w^{pi} y^{pi}$. Appropriate selection of the weights $w$ can specify particular proteins as desirable or repellent. As a problem setting, consider a multi-species RFM network with $M$ species and $N$ ribosomes. There will be $M+1$ pools: $z_E$ for the empty ribosomes and $z_p$ for the translating species. 
Assume that $p=1$ represents the host ribosome with host genes, 
and that each circuit gene may only accept ribosomes from a single pool $z_p$.
The matching of pools and circuit genes that maximizes $y_w$ requires solving a binary assignment problem. If a new synthetic gene was to be designed in an environment where there is severe competition for circuit ribosomes, it may be favorable for this new gene to employ host ribosomes. This can be generalized to multi-species orthogonal ribosomal networks.  
An alternative problem can be formulated in terms of equilbrium constants $K_p$. If there exists sufficient flexibility to adjust $k_p^+ \in [k_p^{+,min}, k_p^{+,max}]>0$ and $k_{p}^- \in [k_{p}^{-,min}, k_{p}^{-,max}]>0$, then: \[
K_p \in [K_p^{min}, K_p^{max}] = \left[\frac{k_p^{+, min}}{k_{p}^{-,max}}, \frac{k_p^{+, max}}{k_{p}^{-,min}}\right] > 0.\]
For each point $K = \{K_p\}_{p = 1}^M$, the weighted protein output $y_w(K)$ can be obtained by finding the steady state with respect to $K$ using Algorithm \ref{alg:rfm_general} and then evaluating $y_w = \sum w^{pi} y^{pi}$. An optimization problem to maximize $y_w$ at steady state can therefore be formulated:
\vspace{-0.04in}
\begin{align}
\label{eq:general_model_opt}
y^*_w &= \max_{K_p} \sum_{p,i} w^{pi} y^{pi}, \\\nonumber
\text{\itshape subject to}:&~ K_p  \in [K_p^{min}, K_p^{max}]. \, 
\mbox{(Dynamics in \eqref{eq:general_model}-\eqref{eq:general_model_}.)}\nonumber
\end{align}

\noindent The search variables are $K$, and the steady states $(z_E, z_p, e^{pi}_j)$ are derived quantities of $K$. 
Generic solvers such as a Grid Search, Bayesian Optimization, and Basin Hopping can be used to approximate $K^*$.
It can be verified that $\partial e^{pi}_j/\partial K_p > 0$ and ${\partial e^{p'i}_j}/{\partial K_p} < 0$ for $p' \neq p$ by evaluating derivatives of steady states. Raising $K_p$ increases $z_p$, $e^{pi}_j$, $y^{pi}$ at the expense of $z_E$ and $z_{p'}$, $e^{p'i}_j$, $y^{p'i}$ for $p' \neq p$. 

\begin{figure}[t]
	\centering
	\begin{subfigure}[b]{0.45\linewidth}
		\includegraphics[width = \textwidth]{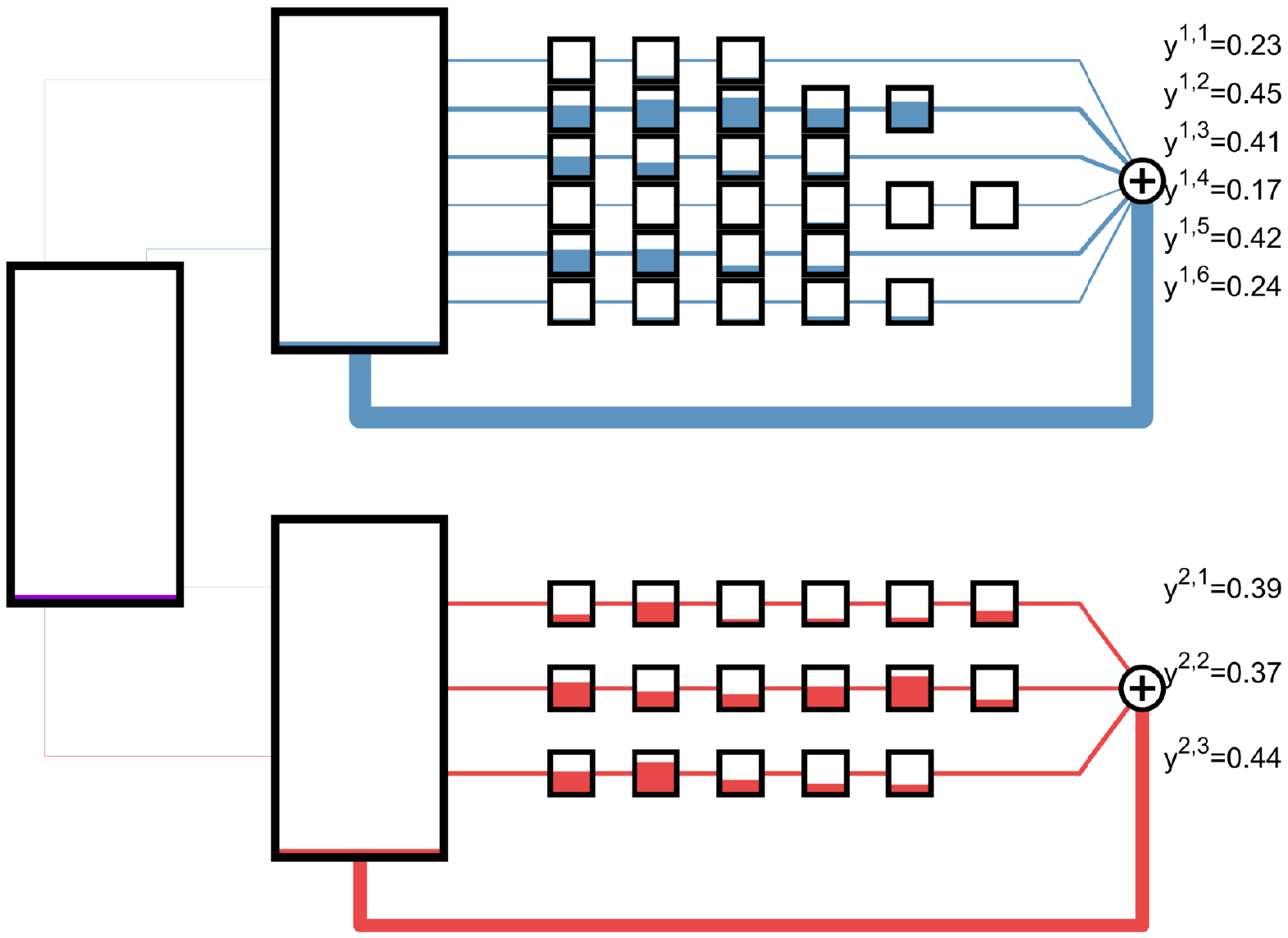}
		\caption{\label{fig:opt_parity} Parity: $y_W=3.11$}
	\end{subfigure}
	\begin{subfigure}[b]{0.45\linewidth}
		\includegraphics[width = \textwidth]{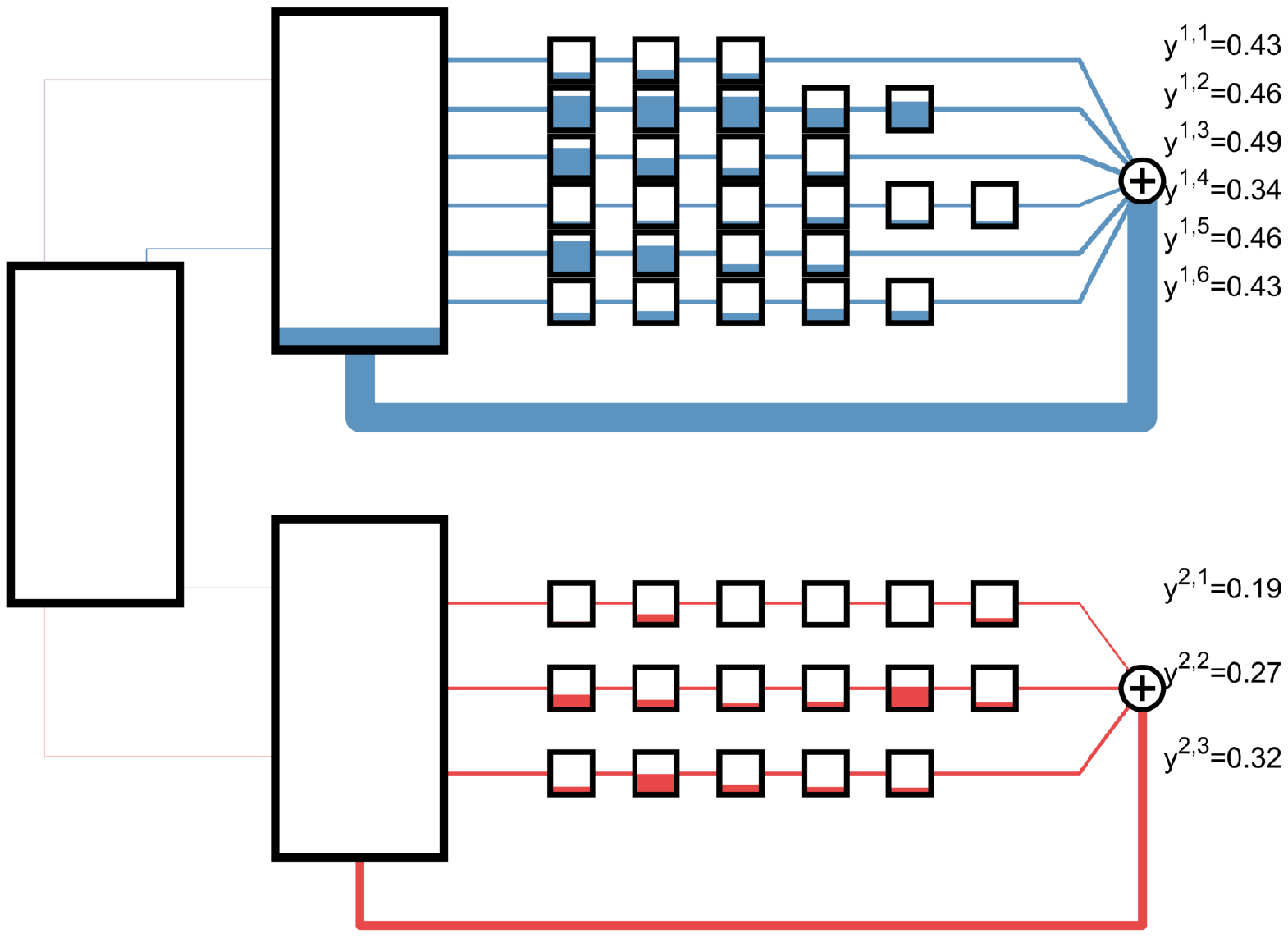}
		\caption{\label{fig:opt_opt} Optimal: $y_W = 3.39$}
	\end{subfigure}
	\caption{\label{fig:optimal_rfm}Optimize the total protein output $(w = \mathbf{1})$}
\end{figure}

\begin{figure}[t]
	\centering
	\includegraphics[width=0.5\linewidth]{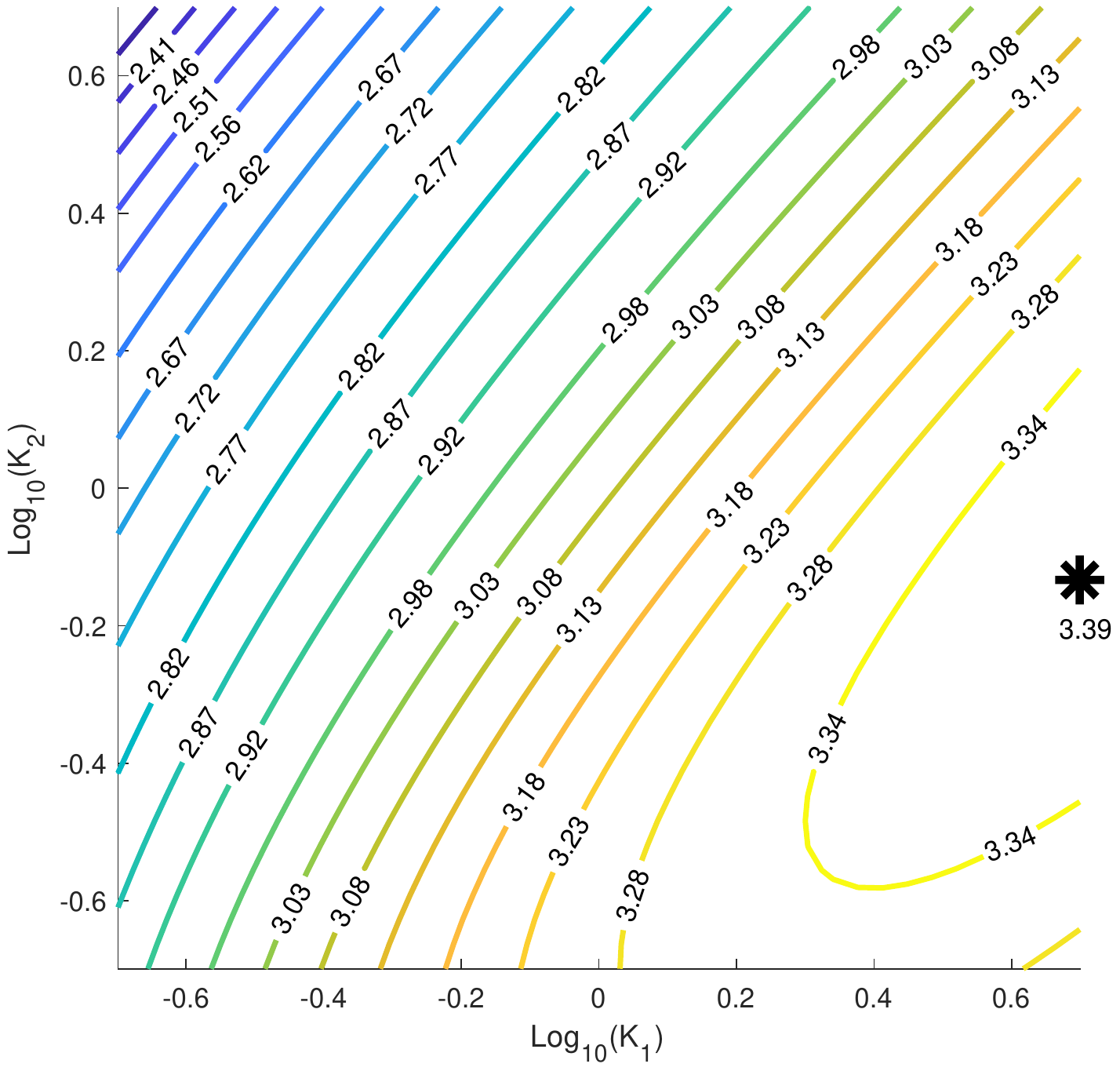}
	\caption{\label{fig:optimal_landscape}Contours of total protein output}
\end{figure}

Fig \ref{fig:optimal_rfm} shows an ORFM with $s_1 = 6$ and $s_2 = 3$ with an objective of maximizing total protein output$(w = \mathbf{1})$. If all mRNA were connected to a common pool of ribosomes (RFMNP), the resultant output is $y_w = 3.18$.  When $K = [5, 5]$, then $y_2 = 3.11$ as shown in Figure \ref{fig:opt_parity}. Solving the optimization problem in Equation \eqref{eq:general_model_opt} with $K^{min} = 1/5$ and $K^{max} = 5$ results in $K^* = [5, 0.734]$ and the protein output is $y_w = 3.38$ in Figure \ref{fig:opt_opt}. 
At steady state in the optimal allocation, $z^H$ will contain $74\%$ of all ribosomes in the pools $(\bar{N}_r)$. The optimization landscape of $y_w$ vs. $(\log_{10}(K_1), \log_{10}(K_2))$ is displayed in Figure \ref{fig:optimal_landscape}.
\vspace{-0.04in}

\section{Conclusion}
\label{sec:conclusion}

Competition for finite resources are inevitable in  protein translation. Orthogonal ribosomes have been developed to boost protein throughput by decoupling circuit genes from the host pool of ribosomes. We extended the existing RFM to orthogonal ribosomes, and generalized the system to an arbitrary number of ribosomal species. Stability results through RLFs and a simple algorithm to compute steady states were shown for the ORFM system. 
A self inhibiting feedback controller can adjust ribosomal production as needed.
Maximizing the weighted sum of protein throughput can be formulated as a low-variable nonconvex optimization problem. Future work includes matching results with lab experiments and allowing for cross-talk in translation.

\section*{Appendix: Proof of Theorem \ref{t.stability}}
\label{sec:appendix_proof}

For simplicity, assume that $s_p=s, n^{p,i}=n$, for all $p,i$. Also, assume $m^{p,i}=1$, for all $p,i$. The argument for the general case will be the similar. Denote $\sigma(x):=\sgn(x)$.

To use the techniques in \cite{MA_TAC,MA_LEARN}, we lift \eqref{eq:general_model}-\eqref{eq:general_model_}  to a higher-dimensional space by defining the \emph{vacancy}   $w_j^{p,i}:=1-x_j^{pi}$, for all $j,i,p$. Hence, terms of the form $(1-x_{j-1}^{pi}) x_j^{pi}$  take the familiar Mass-Action form: $w_{j-1}^{pi} x_j^{pi}$. We will generalize this further by considering arbitrary \emph{monotone} functions of the form $R(w_{j-1}^{pi}, x_j^{pi})$. Hence, we write \eqref{eq:general_model}-\eqref{eq:general_model_} as:
\begin{align}
\nonumber
\dot{z}_E &= \textstyle\sum_p R_p^{-} (z_p) - \textstyle\sum_p R_p^{+} (z_E)  \\ \nonumber
\dot{z}_p &=  R_p^{+} (z_E) - R_p^{-} (z_p)  + \textstyle\sum_i   \left ( R_{n}^{pi}(x_{n}^{pi}) -R_0^{pi} (w_1^{pi},z_p) \right )\nonumber\\   \nonumber
\dot{x}_1^{pi} &= R_0^{pi} (w_1^{pi},z_p) -R_1^{pi}(w_2^{pi}, x_1^{pi})  \\ \label{sys}
\dot{x}_j^{pi} &=R_{j-1}^{pi}  (w_{j}^{pi}, x_{j-1}^{pi} )-R_{j}^{pi}  (w_{j+1}^{pi}, x_{j}^{pi} ) \\ \nonumber
\dot{x}_{n}^{pi} &=R_{n-1}^{pi}  (w_{n}^{pi}, x_{n-1}^{pi} )-R_{n}^{pi}  (  x_{n}^{pi} ).
\end{align}

We only assume that the rates $R_j^{p,i}, R_p^{\pm}$ are monotone w.r.t their reactants, see \cite{MA_LEARN} for the full assumptions. 

Consider $V(x,z)=\sum_{i,p,j} |\dot x_{j}^{pi}|+\sum_p|\dot z_p|+|\dot z_E|$. We  show that $V$ is non-increasing.  Using \eqref{sys}, note that the space can be partitioned into regions, and $V$ is \emph{linear} in \emph{rates} on each region.  Fix an \emph{open} region $\mathcal W$ and write $V=\sum_{p,i,j} \alpha_j^{pi} R_j^{pi}(x,z)+ \sum_p \beta_p (R_p^-(z_p) - R_p^+(z_E))$ on $\mathcal W$. Note $V$ is differentiable on $\mathcal W$, and   the signs of the currents $\dot x_j^{pi}, \dot z_p, \dot z_E$ are constant on $\mathcal W$. We claim that the derivative of each term in $V$ is non-positive. We show first that  $\alpha_j^{pi}\dot R_j^{pi}\le 0$ for all $p,i,j$. We study three cases: $j=0,n$, $0<j<n$. If $j\ne0,n$, then $R_j^{pi}$ appears in   $\dot x_{j}^{pi},  \dot x_{j+1}^{pi}$. If both have the same sign on $\mathcal W$, then \eqref{sys} implies   $\alpha_j^{pi}=0$.  Therefore, $\alpha_j^{pi}\ne 0$ implies $\sigma(\alpha_j^{pi})=-\sigma(\dot x_{j}^{pi})=\sigma(\dot x_{j+1}^{pi})= -\sigma(\dot w_{j+1}^{pi})$. Hence, $\alpha_j^{pi}\dot R_j^{pi}=\alpha_j^{pi}(\frac{\partial R_j^{pi}}{\partial x_j^{pi}} \dot x_j^{pi} + \frac{\partial R_j^{pi}}{\partial w_{j+1}^{pi}} \dot w_{j+1}^{pi}  ) \le 0$ as claimed, where non-negativity of the partial derivatives follows from monotonicity. Next, consider the case $j=0$ where $R_0^{pi}$ appears in two currents $\dot x_{1}^{pi},  \dot z_{p}$. Similar to the previous case, we conclude that if $\alpha_0^{pi} \ne 0$ then $\sigma(\alpha_0^{pi})=- \sigma (z_{p})=-\sigma(\dot w_{1}^{pi})$. Since $R_0^{pi}$ is a monotone function of $w_{1}^{pi},   z_{1}^{pi}$, then $\alpha_0^{pi}\dot R_0^{pi}\le 0$ as claimed. Finally, if $j=n$, similar analysis can be repeated to conclude that if $\alpha_n^{pi}\ne 0$ then $\sigma(\alpha_n)=- \sigma (x_n^{pi})$ and  $\alpha_n\dot R_n^{pi}\le 0$ as claimed.   Next, we want to show that $\beta_p \dot R_p^- \le 0$. Fix $p$. Note that $R_p^-$ appears in two currents: $\dot z_p, \dot z_E$. If they have the same sign then $\beta_p=0$. If they have different signs then \eqref{sys} implies that $\sigma(\beta_p)=-\sigma(\dot z_p)=\sigma(\dot z_E)$. Since $R_p^-$ is a function of $z_p$, then  $\beta_p \dot R_p^- \le 0$. A similar argument can be replicated to show that $-\beta_p \dot R_p^+ \le 0$.  Since $\mathcal W$ and $p,i,j$ were arbitrary, we conclude that $\dot V \le 0$  over the interior of all regions. The boundaries between the regions can be dealt with via the definition of the upper Dini's derivative, see \cite{MA_TAC}.  This concludes the proof of the first statement of Theorem \ref{t.stability}.

We proceed to prove part 2.  The RLF utilized is non-strict, hence it cannot yield Global Asymptotic Stability (GAS) directly. The LaSalle  argument proposed in \cite{MA_TAC} can be used, but it is long. Alternatively, we will show that the Jacobian of \eqref{sys} (reduced to the invariant space for a fixed $N_r$) is robustly non-degenerate. This, coupled with the RLF, implies the GAS of any steady-state \cite{blanchini17,MA_LEARN}. 

To prove non-degeneracy, it has been shown in \cite{MA_cdc14,MA_LEARN} that the Jacobian $J$ of \eqref{sys} at any point in the space can be written as $J=\sum_{\ell=1}^L \rho_\ell Q_\ell$ for some $\rho_\ell\ge 0$, rank-one matrices $Q_\ell$ and some $L$. Here, $L=M(2+s(2n+1))$. The coefficients $\{\rho_\ell\}$  correspond to the partial derivatives of   the rates $\{R_j^{pi}\},\{R_p^{\pm}\}$, while $\{Q_\ell\}$ correspond to the network structure and are independent of the rates. Furthermore, it has been shown also \cite{MA_LEARN} that the existence of an RLF implies that it sufficient to find \emph{one positive point} $\rho^*=(\rho_1^*,..,\rho_L^*)$ for which the (reduced) Jacobian is non-degenerate to show that is non-degenerate for all $\rho_\ell >0$. We will find that point next by studying the structure of the Jacobian. 

Similar to a single pool \cite[SI]{raveh2016model}, it can be easily seen that $J$ has non-negative off-diagonals and strictly negative diagonals (i.e, $J$ is Metzler). In addition, conservation of the number of ribosomes implies that $\mathbf 1^T J=0$. By the definition of the Jacobian, all the entries in each column  contain only the partial derivatives with respect to the state variable associated to the column. Hence, we can choose the corresponding $\rho_\ell$'s such the diagonal entry in each column is scaled to $-1$. Therefore, we consider the Jacobian evaluated at the chosen point $\rho^*$ such that $J^*=P-I$, where $I$ is the identity matrix and $P$ a nonnegative irreducible column-stochastic matrix. By Perron-Frobenius Theorem, $P$ has a maximal eigenvalue 1 with algebraic multiplicity 1. Therefore, $J^*$ has a single eigenvalue at $0$ and  the remaining eigenvalues have strictly negative real-parts. Hence, the reduced Jacobian at $\rho^*$  is non-degenerate. Robust non-degeneracy and GAS follows.

The existence of a steady-state follows from Brouwer's fixed point theorem since \eqref{sys} evolves in a compact space (for a fixed $N_r>0$), and uniqueness follows from non-degeneracy and GAS. The positivity of the steady-state follows from persistence of the ORFM which can be shown graphically by the absence of critical siphons \cite{angeli07}. \strut\hfill $\blacksquare$

\paragraph*{Acknowledgements}

This research was partially funded by NSF grants 1849588 and 1716623. Jared Miller thanks Prof. Mario Sznaier for his support and discussions.

\end{document}